\documentclass{biophys-new}
\usepackage[utf8]{inputenc}
\usepackage{graphicx}
\usepackage[colorlinks,allcolors=cyan!70!black]{hyperref}
\usepackage{lineno}

\usepackage{lipsum}

\title{A mechanism for sarcomere breathing:  volume change and advective flow within the myofilament lattice}
\runningtitle{Biophysical Journal Template} 
\author[a,b]{J.A. Cass}
\author[b,c]{C. D. Williams}
\author[d]{T. C. Irving}
\author[e]{E. Lauga}
\author[b]{S. Malingen}
\author[b,1*]{T. L. Daniel}
\author[f,1*]{S. N. Sponberg}
\runningauthor{Cass et al.} 

\affil[a]{Allen Institute for Cell Science, Seattle, WA 98109, USA}
\affil[b]{Department of Biology, University of Washington, Seattle, WA 98195, USA}
\affil[c]{Applied ML Group, Microsoft CSE, Redmond, Washington, USA}
\affil[d]{BioCAT and CSRRI, Department of Biological Sciences, Illinois Institute of Technology, Chicago, IL 60616, USA}
\affil[e]{Department of Applied Mathematics and Theoretical Physics, University of Cambridge, Wilberforce Road, CB3 0WA Cambridge, UK}
\affil[f]{School of Physics \& School of Biological Sciences, Georgia Institute of Technology, Atlanta, Georgia 30332, USA}

\corrauthor[*]{sponberg@gatech.edu, danielt@uw.edu}

\papertype{Article}

\begin{document}
\begin{frontmatter}

\begin{abstract}
During muscle contraction, myosin motors anchored to thick filaments bind to and slide actin thin filaments. These motors rely on energy derived from ATP, supplied, in part, by diffusion from the sarcoplasm to the interior of the lattice of actin and myosin filaments. The radial spacing of filaments in this lattice may change or remain constant during contraction. If the lattice is isovolumetric, it must expand when the muscle shortens. If, however, the spacing is constant or has a different pattern of axial and radial motion, then the lattice changes volume during contraction, driving fluid motion and assisting in the transport of molecules between the contractile lattice and the surrounding intracellular space. We first create an advective-diffusive-reaction flow model and show that the flow into and out of the sarcomere lattice would be significant in the absence of lattice expansion. Advective transport coupled to diffusion has the potential to substantially enhance metabolite exchange within the crowded sarcomere. Using time-resolved x-ray diffraction of contracting muscle, we next show that the contractile lattice is neither isovolumetric nor constant in spacing. Instead, lattice spacing is time-varying, depends on activation, and can manifest as an effective time-varying Poisson ratio. The resulting fluid flow in the sarcomere lattice of synchronous insect flight muscles is even greater than expected for constant lattice spacing conditions. Lattice spacing depends on a variety of factors that produce radial force, including crossbridges, titin-like molecules, and other structural proteins. Volume change and advective transport varies with the phase of muscle stimulation during periodic contraction but remains significant at all conditions. While varying in magnitude, advective transport will occur in all cases where the sarcomere is not isovolumetric. Akin to ``breathing,'' advective-diffusive transport in sarcomeres is sufficient to promote metabolite exchange and may play a role in the regulation of contraction itself. 
\end{abstract}

\begin{sigstatement}
Muscle operates at the limits of diffusion's ability to provide energy-supplying molecules to the molecular motors that generate force. We show that fluid pumping resulting from the volume changes of muscle's contractile lattice can assist diffusion. High speed x-ray diffraction of contracting muscle shows that the volume change is even greater than expected from classical muscle contraction theory. Muscle contraction itself can promote the exchange of metabolites when energetic requirements are high, a process akin to the role breathing plays in gas exchange. Such multiscale phenomena likely play an underappreciated role in the energetics and force production of nature's most versatile actuators.
\end{sigstatement}
\end{frontmatter}

\section*{Introduction}
Muscle force production is one of the most energy demanding physiological processes in biology. Yet the diffusion of molecules that supply energy in muscle is constrained because the dense lattice of contractile proteins typically has spacings of only tens of nanometers, with additional regulatory and metabolic proteins further crowding the space available within the lattice  \cite{Kushmerick1969,Carlson:2014aa}. Intracellular flow of fluid and concomittant advective transport could assist diffusion, but would require a mechanism to induce flow into and out of the contractile lattice. Muscle cells as a whole can change volume over the course of many seconds or minutes as osmolites build up \cite{Neering1991,Rapp1998}, but the cells are likely isovolumetric on the shorter time scale of periodic contractions (order $10^1$ to $10^2$ milliseconds). However, the environment within the muscle cell is highly anisotropic, consisting of axially repeated sarcomeres that are radially surrounded by sarcoplasmic reticulum and dispersed mitochondria. If the sarcomeres themselves significantly changed volume, then intracellular flow would be likely because fluid would have to move into and out of the contractile machinery. We explore whether muscle contraction itself has the potential to act as an intracellular pumping mechanism, assisting transport of cellular metabolites into and out of the sarcomeres' myofilament lattice from the surrounding intracellular space through flow-mediated advection. Flow-mediated transport within a muscle cell may be especially important for enabling high-frequency and high-power contractions, such as those occurring during flight \cite{Maughan1999,Askew2001,Syme2002,Tu2004cardiac}, sound production \cite{Girgenrath1999,Rome2005}, and cardiac function \cite{Powers2018, Gordon2000}, where the strain is periodic and energy demand can be high.

In muscle, contractile proteins form a regular lattice composed of myosin-containing thick filaments and actin-containing thin filaments \cite{Irving2007}. The lattice is subtended by z-disks on either end, forming the sarcomere, the fundamental unit of muscle contraction. Each muscle cell contains many sarcomeres that share the same intracellular fluid and require the exchange of metabolites with surrounding organelles. To power the sliding of the thick filaments relative to the thin filaments, each of the billions of myosin motors requires energy derived from ATP hydrolysis. However, the dense packing of the myofilament lattice is presumed to limit the diffusion of critical energetic metabolites (e.g., creatine phosphate, ADP, and ATP) and regulatory molecules (e.g. Ca$^{2+}$) \cite{Kushmerick1969,Carlson:2014aa}. Interestingly, as the thin filaments slide during contraction, and the z-disks to which they attach are pulled toward the midline of the sarcomere, mass conservation demands that either 1) fluid within the sarcomere is squeezed out of the lattice into the intracellular spaces surrounding the sarcomeres, or 2) the filament lattice expands radially, mitigating mass flux by conserving the lattice volume. It is also possible that the \textit{in vivo} dynamics of the lattice follow neither of these limiting cases. Indeed, x-ray diffraction studies point to possible radial motions of the lattice that might influence fluid exchange \cite{Cecchi:1990aa,Cecchi:1991aa,Iwamoto2013,George:2013aa}. Radial motion during natural contractions may result in conditions that are neither constant volume nor constant spacing because electrostatic and structural forces as well as active cross-bridges can modulate radial filament motion \cite{Smith2014,Williams:2010aa,Williams:2012aa}. The interaction between fluid exchange, radial lattice motions and substrate delivery remains unresolved.

\begin{figure*}
\centering
\includegraphics[width= 1\linewidth]{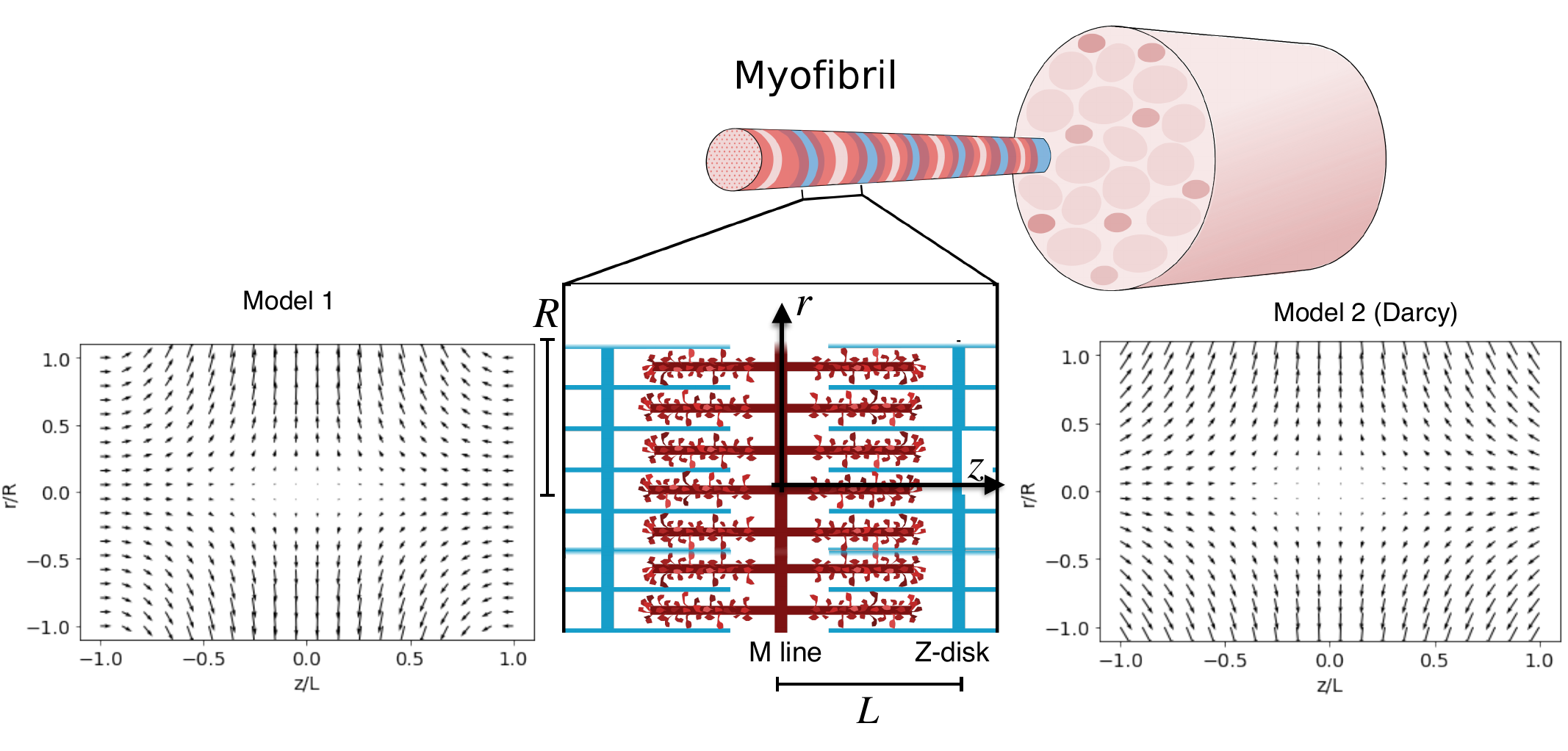}
\internallinenumbers
\caption{ \textbf{Two flow fields for a contracting sarcomere.}  Schematic of contracting and lengthening sarcomeres, with red and blue rods representing thick and thin filaments respectively.  A cross-sectional view of the internal flow field in normalized coordinates during contraction at maximum velocity for the kinematic-based flow field (Model 1, left) and the Darcy flow field (Model 2, right).  Both flow fields have a large radial component at the M-line of the sarcomere ($z$ = 0).  The peak z-disk velocity is -0.0075.}  

\label{fig:fig1}
\end{figure*}

To assess the flow resulting from  lattice volume change during cyclic contractions, we first consider the case of a contracting sarcomere with constant lattice spacing, as observed in the small-strain contractions of \textit{Drosophila} indirect flight muscle \cite{Dickinson:2005aa}. While our simple models do not account for the complexity of myriad interacting sarcomeres and surrounding organelles, they provide a first glimpse into the understudied problem of intrasarcomeric flows. We use these flows to develop a simplified model of how advection, coupled with diffusion and reaction terms,  influences substrate delivery into the densely packed space within the myofilament lattice (Fig. \ref{fig:fig1}A). With diffusion greatly reduced in the crowded subcellular environment of a sarcomere \cite{Carlson:2014aa}, mechanisms that enhance substrate replenishment may have a profound impact on the energetics of muscle contraction. To complement the model of advective transport of substrate delivery, we use time-resolved x-ray diffraction to measure nanometer-scale changes in the radial spacing of the filament lattice of intact muscle contracting under \textit{in vivo} conditions with controlled muscle length and electrical activation.   Measuring the temporal pattern of radial lattice spacing changes together with controlled muscle length and activation allows us to explore volume changes at the level of sarcomeres for \textit{in vivo} muscle contraction. 

\section*{Materials and Methods}

To our knowledge, there are no explicit studies of flow within the sarcomere and fluid exchange between the myofibril and the surrounding intracellular environment.  Here we are concerned with understanding the implications of flow for substrate supply in the sarcomere. While it is intuitive to expect that flow enhances substrate exchange over diffusion alone, its potential contribution remains to be explored for any model of fluid flow in the sarcomere. To explore how the flow induced by contraction affects substrate delivery, we numerically solved the one-dimensional advection-diffusion-reaction equation for ATP supply at the midline (M-line) of the sarcomere:

\begin{equation}
\label{DCR}
\frac{\partial{c}}{\partial{t}} = D \mathbf{\nabla}^2 c - \mathbf{u} \cdot \mathbf{\nabla}c - \chi,
\end{equation} 
\\
where $c$ is the radially-dependent concentration of substrate (ATP), $D$ is the diffusion coefficient taken from measurements of rabbit skeletal muscle fibers \cite{Carlson:2014aa} ($8 \times 10^{-9} \: cm^2 s^{-1}$), $\mathbf{u}$ is the intra-sarcomeric vector flow field, and $\chi$ is the rate of ATP consumption which we have modeled with Michaelis-Menten kinetics.     Both $\mathbf{u}$  and $c$ vary both spatially and temporally.  

\subsection*{Modeling:  Derivation of flow fields within the sarcomere}
Equation \ref{DCR} requires that we specify the flow field that arises from sarcomere length changes for contracting sarcomeres of constant radius. In the absence of a detailed analytic solution to flow in the complex domain of myofilament overlap, or a full computational fluid dynamics simulation of sarcomere flows, we considered two different simple flow fields to explore the general consequences of fluid exchange to substrate delivery. 

In both models, we use a cylindrical model of the half sarcomere in which we consider axial ($z$) and radial ($r$) velocity components in the domain $0 \leq z \leq L$ and  $0 \leq r \leq R$
\begin{equation}
        \vec{u}(r, z) = u_r(r, z)\hat{r} + u_z(r, z)\hat{z}.
    \end{equation}

We next consider two approaches to calculate the relevant flow fields, both of which are based on an effective plug flow model in which $u_z$ varies only in the $z$-direction (the velocity component perpendicular to the z-disk is uniform in the $r$-direction).  In addition, at the z-disk ($z=L$), the fluid velocity is equal to the instantaneous velocity of the sarcomere, $U(t)$, which in our model is periodic and $U(t)$ is positive during lengthening. Hereafter, for simplicity, we refer to $U(t)$ as $U$. All velocities linearly depend on $U$. Additionally, at the M-line ($z=0$) symmetry demands that $u_z = 0$.  Symmetry also demands that $u_r = 0$ at the axial center of the sarcomere ($r = 0$). Thus: 

\begin{subequations}
\label{BC1}
\begin{align}
u_z(r,L) = -U,  \\
u_z(r,0) = 0,\\
u_r(0,z) = 0.
\end{align}
\end{subequations}

Additionally, we assume the flow to be cylindrically symmetric. Both flow fields satisfy the equation of continuity for flow within the sarcomere in cylindrical coordinates $r,z$:
\begin{equation}
\label{con}
 \frac{1}{r} \frac{\partial}{\partial r}(r u_r) + \frac{\partial u_z}{\partial z} = 0
\end{equation}

\subsubsection*{Flow field model 1: no-slip condition at the z-disk}

In addition to the boundary conditions above, our first model assumes the no-slip condition at the z-disk.  In addition, we  sought a flow field that generates no net axial shear at the M-line of the sarcomere (the axial plane of symmetry), no net radial shear along the axis $r=0$, and satisfies continuity.  Thus, in addition to Equation \ref{con}, we have:
\begin{subequations}
\label{BC2}
\begin{align}
u_r(r,L) = 0,\\
\frac{\partial u_r(0,z)}{\partial z}  = 0,\\
\frac{\partial u_z(r,0)}{\partial z}  = 0.\\
\end{align}
\end{subequations}

An admissible flow field that satisfies equations \ref{BC1} - \ref{BC2} is:  
 \begin{subequations}
 \label{FF1}
 \begin{align}
        u_r(r, z) = -U \frac{3}{4}\frac{r}{L} \bigg[ 1 - \Big(\frac{z}{L}\Big)^2 \bigg], \\
        u_z(z) = -U \frac{3}{2} \frac{z}{L}\bigg[ 1 - \frac{1}{3}\Big(\frac{z}{L}\Big)^2 \bigg],
\end{align}
\end{subequations}
\noindent
where ${L}$ is the axial position of the z-disk and all velocities are time dependent and periodic as noted above. 


This flow field generates a peak radial velocity at the M-line of $u_R = 3UR/4L$. If the length of the half sarcomere and its radius are of a similar order of magnitude, this peak velocity will be close to the shortening velocity of the sarcomere. 


\subsubsection*{Flow field model 2: Darcy flow with asymmetric permeability}
The above flow field does not explicitly account for the influence of filaments.  We thus developed a second approach that draws on a Darcy flow (Model 2) in the lattice of filaments which includes radial and longitudinal permeabilities.  



The coordinate system, plug flow assumption, and boundary conditions in equation \ref{BC1} all apply to this model as well.  Here, the plug flow velocity in overlapping filaments is given by 
\begin{equation}
\label{plug}
u_z = -\frac{U}{2} - \frac{k_l}{\mu}\frac{\mathrm{d}p(z)}{\mathrm{d}z},
\end{equation}
where $p(z)$ is the pressure, $\mu$ is fluid viscosity, and $k_l$ is the axial Darcy permeability. 

Conservation of mass also indicates that the axial flow generated by the contracting z-disk motion is balanced by a radial efflux along the sarcomere:
\begin{equation}
\label{Conserv}
\frac{\mathrm{d}Q}{\mathrm{d}z} +2 \pi R u_r(R,z) = 0,
\end{equation}
where $Q(z) = \pi R^2 u_z$.  This efflux is modeled with a second Darcy relationship for the radial border of the sarcomere: 

\begin{equation}
\label{radialDarcy}
u_r(R,z) = \frac{k_r}{R \mu}p(z),
\end{equation}
where $k_r$ is the radial Darcy permeability.  

Equations \ref{plug} - \ref{radialDarcy} lead to a second-order differential equation for the pressure: 
\begin{subequations}
\label{ODE}
\begin{align}
\frac{\mathrm{d^2}p(z)}{\mathrm{d}z^2} = \frac{\alpha^2}{R^2}p(z),\\
\alpha^2 = \frac{2 k_r}{k_l},
\end{align}
\end{subequations}
\noindent
both pf whose general solution is
\begin{equation}
\label{gen}
p(z) = p_1 \mathrm{cosh}\left(\alpha \frac{z}{R}\right) + p_2 \mathrm{sinh}\left(\alpha \frac{z}{R}\right).
\end{equation}

The constants $p_1$ and $p_2$ are found by using boundary conditions from \ref{BC1} and the plug flow model in \ref{plug} to give the flow rate ($Q$) to be $-U \pi R^2$ at $z=L$ and 0 at $z = 0$ :

\begin{subequations}
\label{BC3}
\begin{align}
    \frac{\mathrm{d}p(L)}{\mathrm{d}z} = \frac{\mu U}{2 k_l},\\
   \frac{\mathrm{d}p(0)}{\mathrm{d}z} = \frac{-\mu U}{2 k_l}.
\end{align}
\end{subequations}

These boundary conditions in equation \ref{BC3} along with equation \ref{gen} gives a solution for the pressure and radial efflux velocity: 

\begin{subequations}
\label{uRz}
\begin{align}
p(z) = \frac{\mu U R}{2 k_l \alpha} \frac{\mathrm{cosh}(\alpha \frac{z - L/2}{R})}{\mathrm{sinh}(\alpha \frac{L}{2R})},\\
u_r(R,z) = \frac{U \alpha}{4} \frac{\mathrm{cosh}(\alpha \frac{z - L/2}{R})}{\mathrm{sinh}(\alpha \frac{L}{2R})}.
\end{align}
\end{subequations}

The radial flow within the sarcomere $u_r(r,z) $ can be obtained from  the conservation of mass in equations \ref{con} and \ref{BC2}, and the axial flow velocity in equation \ref{plug} gives 
\begin{subequations}
\label{u(r,z)}
\begin{align}
   \frac{\partial}{\partial r}(r u_r) = -r \frac{\partial u_z}{\partial z},\\
   \frac{}{}
   \frac{\mathrm{d}u_z}{\mathrm{d}z} = \frac{-2 k_r}{\mu R^2}p(z).
\end{align}
\end{subequations}

From the above, by combining equations \ref{u(r,z)} and \ref{uRz} we finally arrive at an expression for the radial flow within the sarcomere:

\begin{equation}
    u_r(r,z) = \frac{r}{R}u_r(R,z).
\end{equation}

Finally, we use equations \ref{plug}, \ref{radialDarcy} and \ref{uRz} to arrive at the axial (z-component) of the flow field:

\begin{equation}
    u_z(z) =\frac{-U}{2} \left(1+\frac{\mathrm{sinh}(\alpha \frac{z - L/2}{R})}{\mathrm{sinh}(\alpha \frac{L}{2R})}\right).
\end{equation}

The flow field that results from this analytic solution shows that both the radial and axial velocities are maximum at the perimeter of the sarcomere nearest the z-disk.  This contrasts with the prior flow field that has the velocity maximum at the perimeter of the sarcomere nearest the M-line.  Note that the Darcy model does not satisfy the no-slip condition at the z-disk ($z = L$).

Flow fields for both models are shown in Figure 1.  Both are symmetric about the M-line, and both show a radial velocity component that varies linearly with $r$.

\subsubsection*{Substrate delivery: computing concentration}
We solve for $c$ in equation \ref{DCR} for a cylindrical sarcomere 1.5 $\mu$m in radius and 3 $\mu$m long undergoing periodic length changes of 0.15 $\mu$m (5\%) amplitude at 25 Hz, consistent with synchronous insect flight muscle with a high metabolic power requirement \cite{George:2013aa,Tu2004cardiac}.   We consider two comparisons: one in which diffusion and advection contributes to substrate exchange in the absence of any consumption ($\chi= 0$) and the other in which the ATP consumption rate follows Michaelis-Menten kinetics:
\begin{equation}
\label{MichMent}
    \chi = \frac{V_m [ATP]}{K_m + [ATP]},
\end{equation}
\noindent
for which Mijailovich \textit{et al.} \cite{Mijailovich2017}   provided estimates of $V_m$ to be 29.3 mM/s and $K_m$ to be 101 mM.   In both cases, we hold the extracellular concentration of ATP to be 10 mM. We also calculate $c$ for sarcomeres of difference size. 

\subsection*{Modeling:  Numerical solution of the advection-diffusion equation}  

We developed a Python-based numerical PDE-solver to calculate the concentration of ATP in the sarcomere as a function of time and radial position \cite{Oliphant:2006aa}. We solved the PDE for the initial condition of 0 mM internal ATP concentration at time 0 with a boundary condition of 10 mM ATP at the cylinder radius. We calculated the concentration by solving the one-dimensional diffusion-advection equation (Eq 1), with the time dependent flow field described in the previous section (Eq 2-4) serving as $\mathbf{u}$. We discretized this equation using a central-difference Euler finite differencing scheme, and solved it along a radial grid at the axial center of the sarcomere cylinder. This equation was then sequentially solved in steps along a temporal grid; the resolution of the temporal and radial grids were coupled by a von Neumann stabilization criterion:
    
    \begin{equation}
        \Delta t<\frac{(\Delta r)^{2}}{2D},
    \end{equation}
where $\Delta t$ and $\Delta r$ are the temporal and radial grid resolutions, and $D$ is the diffusion coefficient. For our model, we prescribe:
 $\Delta t = 0.9 (\Delta r)^2/(2D)$, where the $0.9$ factor ensures the criterion is met. The result of this simulation is a numerical calculation of the ATP concentration on a radial-temporal grid, during many cycles of successive sarcomere contractions.\\

\subsection*{Experiment:  Preparation of specimens} Hawkmoths (\textit{Manduca sexta}) were grown at the University of Washington Insect Husbandry Facility.  Moths were cold anesthetized at 4$^{\circ}$C with a thermo-electrically cooled stage after which wings, head, and legs were removed. 
The left and right pair of dorsolongitudinal downstroke muscles (DLMs) were isolated and together mounted in the apparatus as described previously \cite{Tu2004submax, George:2013aa}. The anterior phragma region of the scutum, where the DLMs originate, was rigidly adhered to a custom brass mount shaped to the curvature of the thorax. The DLMs insert onto the second phragma, an invagination of the exoskeletal between the meso- and meta-thoracic segments. A second brass mount with two stainless steel prongs was inserted into the phragma to provide a rigid attachment. This mount connected to a muscle ergometer (Aurora Scientific 305C) that controlled muscle axial length and measured force. After mounting, the ventral side of the thorax was removed immediately below the DLMs to sever the upstroke dorsoventral muscles and the steering muscles. A $\sim$3~mm strip of exoskeleton was then removed circumferentially around the thorax to mechanically release the DLMs from the thorax. This procedure results in the pair of isolated DLMs mounted between two sections of the thorax (the first phragma at the anterior of the scutum and the posterior phragma). The muscle was set to its \textit{in vivo} operating length, $L_{op}$, which is $0.98L_{rest}$ \cite{Tu2004submax}. Two bipolar tungsten-silver wire electrodes were inserted through the five subunits of each DLM. Stimulation amplitude (bipolar potentials, 0.5 ms wide) was set by monitoring the isometric twitch response in the muscle and setting the stimulation voltage to the twitch threshold plus 1 V. Typically 3 V stimuli were used.\\

\subsection*{Experiment:  Time-resolved small angle x-ray diffraction} X-ray experiments used the small angle instrument on the BioCAT undulator-based beamline 18-ID at the Advanced Photon Source (Argonne National Laboratory, Argonne, Illinois). The overall experimental arrangement is shown in Fig. \ref{fig:fig3}. The x-ray beam energy was 12 keV (wavelength 0.103~nm), with a specimen-to-detector distance of 3.2 m. Fiber diffraction patterns were recorded with a photon-counting Pilatus 100k (Dectris Inc.) pixel array detector collecting images at 125 Hz. A rapid shutter closed during the 4 ms detector readout time. Beam intensity was adjusted with aluminum attenuators to be in the range $10^{11}$ -$10^{13}$ photons s$^{-1}$, which provided adequate counting statistics in the x-ray pattern with minimal radiation damage to the specimen. In addition, the preparation and the entire experimental apparatus was oscillated in the beam with a large stepper ($\sim$1 mm amplitude, $\sim$10 Hz) to reduce the x-ray dose on a given region of the muscle. X-ray images were processed using an automated machine vision algorithm \cite{Williams:2016} that fits both equatorial diffraction intensity peaks and a diffuse background scattering profile. Only trials in which the program resolved diffraction peaks were used in analysis. The distance between the first diffraction peaks, $S_{10}$, is related to the inter-filament lattice spacing, $d_{10}$, by Bragg's law. Sample-to-detector distance was calibrated using a silver behenate scattering image. 

Physiologically realistic patterns of periodic axial strain and stimulation were applied to the whole muscle using the ergometer and stimulator \cite{Josephson1985, Tu2004submax}. For most trials, we recorded 100 cycles and images were collected at the same phase within each cycle. The first 10 cycles were not considered so that data were collected entirely under steady periodic conditions. Resulting $d_{10}$ measurements were averaged at each phase over the remaining 90 cycles. We used a whole, isolated muscle preparation to best recreate \textit{in vivo} conditions that occur during actual locomotion and report percent change relative to the mean across all preparations. Strain can be heterogeneous across the sarcomeres of an intact muscle undergoing contractions. While diffraction imaging averages over a small volume of sarcomeres (the beam was $\sim 30 \times 150$ $\mu{m}^2$), regional strain heterogeneity as well as variable degrees of local heating or damage due to x-ray imaging likely contribute to these different mean values.  Because of this variation, we centered $d_{10}$ around each preparation's mean before calculating the percent change.\\

\section*{Results and Discussion}

\subsection*{Flow assisted transport can enhance substrate delivery} 

\begin{figure*}
\centering
\includegraphics[width= 1\linewidth]{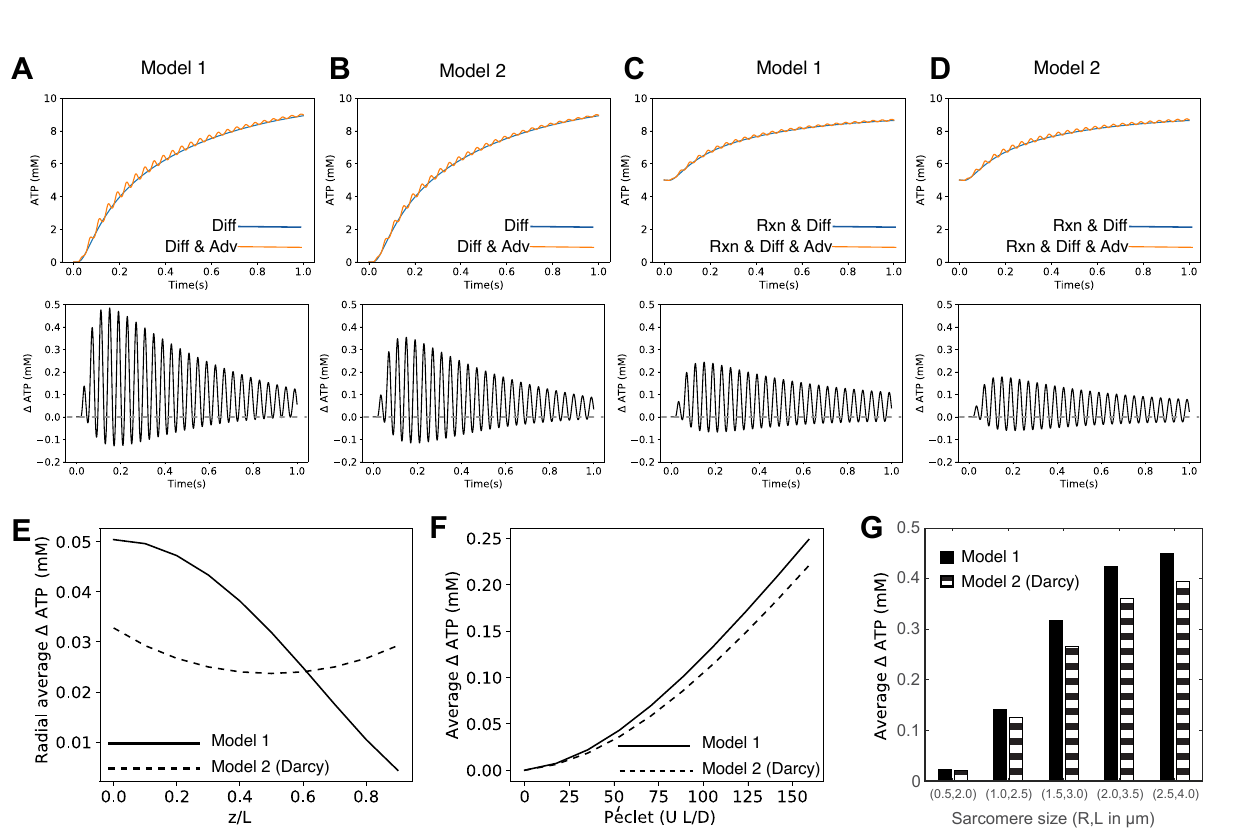}
\internallinenumbers
\caption{ \textbf{Substrate delivery for the two flow fields with and without reaction.}  Here we plot the time history of concentration changes at the M-line at a radial point R/3.  The upper panels show the time history of substrate delivery by diffusion alone (blue lines) and via both diffusion and advection (orange lines). Beneath the upper row are plots of the difference between substrate delivery by diffusion alone and that by both diffusion and advection. Note that the mean of the oscillations (the cycle averaged advective advantage) is always grater than zero (dashed grey line). Panels A and B correspond to concentration changes for a muscle fiber that is initially devoid of ATP for flow model 1 (A) the flow model 2 (B).  Panels C (flow model 1) and D (flow model 2) plot the concentration changes for a muscle fiber with an  initial concentration of 5 mM ATP and an ATP consumption rate that follows Michaelis-Menten kinetics.   Both flow fields yield similar results for simulations with and without reaction kinetics. Panel E shows the cycle-averaged advective advantage as a function of radial position for each of the two models for the diffusion-advection-reaction cases (C, D). Panel F shows that the advective advantage increases with the P\'eclet number, the ratio of flow-driven to diffusive transport. Panel G shows that advective advantage increases with size of the sarcomere (see also Supplemental Figure 1.}
\label{fig:fig2}
\end{figure*}

We computed the flow field by satisfying the appropriate boundary conditions for the half-sarcomere. The resulting flow is symmetric (Fig. \ref{fig:fig1}); with the addition of permeability, the field remains similar in shape, but changes in magnitude. The resultant vector flow field is significant at the scale of the sarcomere and evident in two time-lapse visualizations, provided in the Supplemental Materials, (i) a 2-D slice through the cylinder interior and (ii) the 3-D flow through the cylinder surface (Movie S1 and S2 in the Supporting Material, respectively). With an initial substrate concentration difference of 10 mM between the sarcomere exterior and interior, we find that advection augments the delivery of ATP into the sarcomere over diffusion alone (Fig. \ref{fig:fig2}A-D). The difference in ATP delivery between these two models (advection-diffusion and diffusion-only) oscillates as advection pumps ATP in and out during cyclical contraction, but the average difference over each cycle and summed across the radius is always positive (Movie S3 in the Supporting Material). 

We found that this advective advantage holds true even in conditions where ATP is consumed by crossbridge cycling. To demonstrate this, we next added the Michaelis-Menten reaction term for the two flow fields with an initial 5 mM uniform concentration of ATP and a fixed 10 mM concentration of ATP outside the sarcomere (Fig. \ref{fig:fig2}C,D). We cycle-averaged the difference in the radially-averaged concentrations from each of these models to obtain an overall estimate of advective advantage (Fig. \ref{fig:fig2}E). Both flow fields have a maximal advective advantage at the M-line.  Additionally, while both flow fields lead to a similar overall cycle-averaged advective advantage, the Darcy model (model 2) of flow is axially more uniform than the kinematic model of flow (model 1).  
Both the magnitude of the shortening velocity and diffusion coefficient influence the cycle-average advective advantage. The P\'eclet number ($Pe = UL/D$) measures the influence of these two factors. As diffusion becomes more limiting, advection plays an ever greater role (Fig. \ref{fig:fig2}F). We find that higher contraction velocities (higher frequency or larger amplitude) will drive greater rates of substrate exchange. Correspondingly, different molecules will have different diffusion coefficients depending on size. The effect of a lower diffusion coefficient can also be captured by a higher P\'eclet number. High frequency, large amplitude contractions, like those of synchronous insect flight muscle, would have enhanced advective advantage, especially when they are concomitant with high power requirements.

The flow fields we have used in this study are approximations to the real flow that result from the complex geometry and motions of a sarcomere. Our goal in using these estimates of flow was to explore how bulk motion may influence substrate delivery and to establish a basis on which future research could more deeply investigate this unexplored aspect of intra-sarcomeric transport.   We are, however, acutely aware that myriad factors may influence the flow, including more complex lattice motions (changing sarcomere radius) and flows between radially and axially adjacent sarcomeres. As a simple example, advective advantage increases in sarcomeres with larger volumes, showing that the impact on intracellular transport likely varies across system (Fig. \ref{fig:fig2}G). Additionally, the porosity of the z-disk might allow axial flow, challenging our assumption of the no-slip condition for Model 1. Finally, while sarcomeres can be tightly packed, recent reconstructions show that myofibrils form a branching network that can have significant spacing between sarcomeres \cite{Willingham2020}. Neighboring sarcomeres may change boundary conditions and influence the resulting flow, but as long volume change occurs, convection is necessary and may actually occur at the scale of the whole myofibril network. Addressing such issues would likely rely on computational fluid dynamics in which both complex motions and complex geometry can be specified. A first, recent finite element model of a sarcomere found that the explicit flow field is significant and radial flow was maximal at the free ends of the myofilaments \cite{Malingen2021}.


The advantage of "pumping" via sarcomeric flow would apply not only to ATP, but also to any appropriately sized substrate with lower concentration inside the lattice than outside. Conversely, molecules that are more concentrated within the lattice (\textit{e.g.}, ADP) would benefit from flow-assisted transport out of the sarcomere. The advantage is greatest during the earliest cycles of the simulation (Fig. \ref{fig:fig2}A-D). As time elapses and the cylinder fills with ATP, the advantage afforded by advection diminishes. It is interesting to note that this decrease is not monotonic. While the diffusion and diffusion-advection models must eventually equilibrate (Fig. \ref{fig:fig2}A-B), the constant sink introduced by reaction kinetics within the sarcomere produces a more persistent and steady advective advantage (Fig. \ref{fig:fig2}C-D), which only appears smaller because of the differenc in initial conditions (5 mM vs. 10 mM concentration difference). Depletion of a substrate enhances the advantage of \textit{in vivo} advection-assisted delivery within the sarcomere.

Other processes, such as the creatine phosphate pathway, also provide alternative ways of regenerating ATP. The overall implications specifically for ATP balance would need to take into account the various source and sink pathways for ATP metabolism. However, differences in the concentration of ATP do exist across the cell and within the sarcomere and so advection will enhance ATP delivery, contribute to transport of other metabolites, and influence reaction processes. It is the magnitude of the effect that will vary with the particular cellular environment considered. While we model the effects on ATP as an example, intrasarcomeric flow could be significant for other molecules, especially those involved in regulation, like Ca$^{+2}$, or fatigue, like inorganic phosphate.


\subsection*{Lattice spacing change within synchronous insect flight muscle is neither constant nor isovolumetric} 
While constant lattice conditions would lead to significant intra-sarcomeric flows and increases in substrate delivery, it is reasonable to ask if radial motions of the lattice during natural conditions reduce or augment such volume changes. Indeed, it is possible for the radial spacing to change in such a way that would lead to a constant volume for the lattice: as the z-disks move inward, the radial spacing would increase inversely with the square root of the sarcomere length, corresponding to a Poisson ratio of 0.5. Time-resolved x-ray diffraction allows measurement of the radial myofilament lattice spacing under physiological conditions \cite{Irving2007,Iwamoto2013,Dickinson:2005aa}. Prior analyses have led to disparate interpretations concerning volume changes of the sarcomere during contraction. Early evidence supported the constant lattice volume hypothesis \cite{Huxley:1953aa,Matsubara:1972aa}, but more recent evidence suggests otherwise:  in the asynchronous flight muscle of \textit{Drosophila}, the radial spacing of the lattice did not vary down to angstrom resolution \cite{Irving:2000aa,Dickinson:2005aa}. However, this muscle has a very small $\sim1 \%$ axial strain during contraction. In tetanically activated frog muscle fibers, the radial spacing of the lattice changes during rapid length perturbations, but in a way that yields measurable volume changes of the lattice \cite{Cecchi:1990aa,Cecchi:1991aa}. In contrast to the results from \textit{Drosophila}, time-resolved x-ray studies of a cardiac-like synchronous insect flight muscle in \textit{Manduca sexta} showed significant changes in the radial spacing of the lattice during cyclic contractions, with the temporal pattern of radial changes influenced by the timing of cross-bridge activation \cite{George:2013aa}. However, the contraction conditions used in that study were reduced in their axial strain. The question of how changes in lattice spacing map to sarcomere volume has not been resolved for the \textit{in vivo} function of synchronous insect flight muscle, which has high power requirements.

\begin{figure*}
\centering
\includegraphics[width=.8\linewidth]{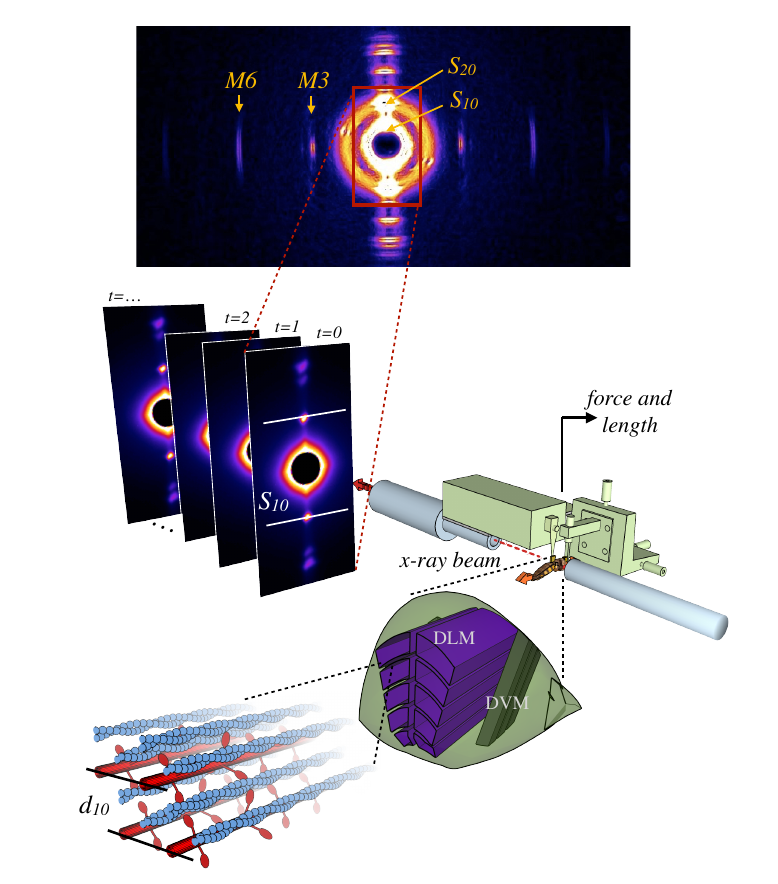}
\internallinenumbers
\caption{ \textbf{Schematic of x-ray fiber diffraction procedure for measuring lattice spacing.} 
From bottom left: the lattice of contractile proteins, containing myosin (red) and actin (blue) filaments is shown, with the lattice spacing ($d_{10}$) marked. Zooming out, these parallel filaments are part of the dorsolongitudinal muscle (DLM, purple) which runs perpendicularly to the dorsoventral muscle (DVM, dark green) in the hawkmoth thorax (light green). The DVMs and a strip of exoskeleton are removed and the mechanically isolated DLM muscles are placed in the synchrotron x-ray diffraction beam (dotted red line). Muscle length is controlled and the force is measured by a force-feedback motor (light green). As the beam line passes through the sample, the resulting diffraction pattern is recorded on a detector. This is repeated over the course of many contractions, resulting in a series of frames of the x-ray diffraction pattern. On each pattern, two bright spots (marked $S_{10}$) are used to calculate the $d_{10}$ interfilament lattice spacing. }
\label{fig:fig3}
\end{figure*}

Here we combine controlled length and activation of muscle with simultaneous, time resolved x-ray diffraction to show that there are significant changes in both the radial spacing of the filament lattice and the volume of the lattice. We do so using a model preparation (\textit{M. sexta}) under physiologically relevant length change and activation conditions. Like most flying insects, hawkmoths power their flight by two dominant muscle groups: the dorsolongitudinal muscles which drive downstrokes of the wings and the dorsoventral muscles which drive upstrokes. We sinusoidally oscillate the moth DLMs at 25 Hz (a physiologically relevant frequency) under controlled timing of the phase of activation with simultaneous time-resolved imaging from x-ray diffraction (see Methods). We measured the distance between the interfilament planes (termed $d_{10}$) from the separation of the first x-ray diffraction peaks ($S_{10}$) with millisecond resolution to provide high-speed imaging of the radial motions of the myofilament lattice during the contraction cycle (Fig. \ref{fig:fig3}).

We also tested the effect of different patterns of lattice spacing. In the specialized, asynchronous flight muscle of \textit{Drosophila}, contraction is decoupled from neural activation. In contrast, \textit{M. sexta} powers wingstrokes with muscles that are activated synchronously via motor neuron control \cite{Tu2004submax}. Altering the phase of stimulation influences the power output of the muscles \cite{Tu2004submax}, and can also influence the time course of lattice spacing change. Using a whole, but isolated, muscle preparation allows us to precisely and repeatedly specify the timing of electrical stimulation and produces less variable x-ray diffraction patterns than seen in whole-animal \textit{in vivo} preparations\cite{Malingen2020}. By changing the phase of electrical stimulation, we can influence crossbridge binding and hence radial force production \cite{George:2013aa,Sponberg2012}. However, while this will likely produce variation in the pattern of lattice spacing, it is not the only determinant. Other sources of radial force production will influence the particular patterns of radial lattice spacing that emerge.

\begin{figure*}
\centering
\includegraphics[width=.7\linewidth]{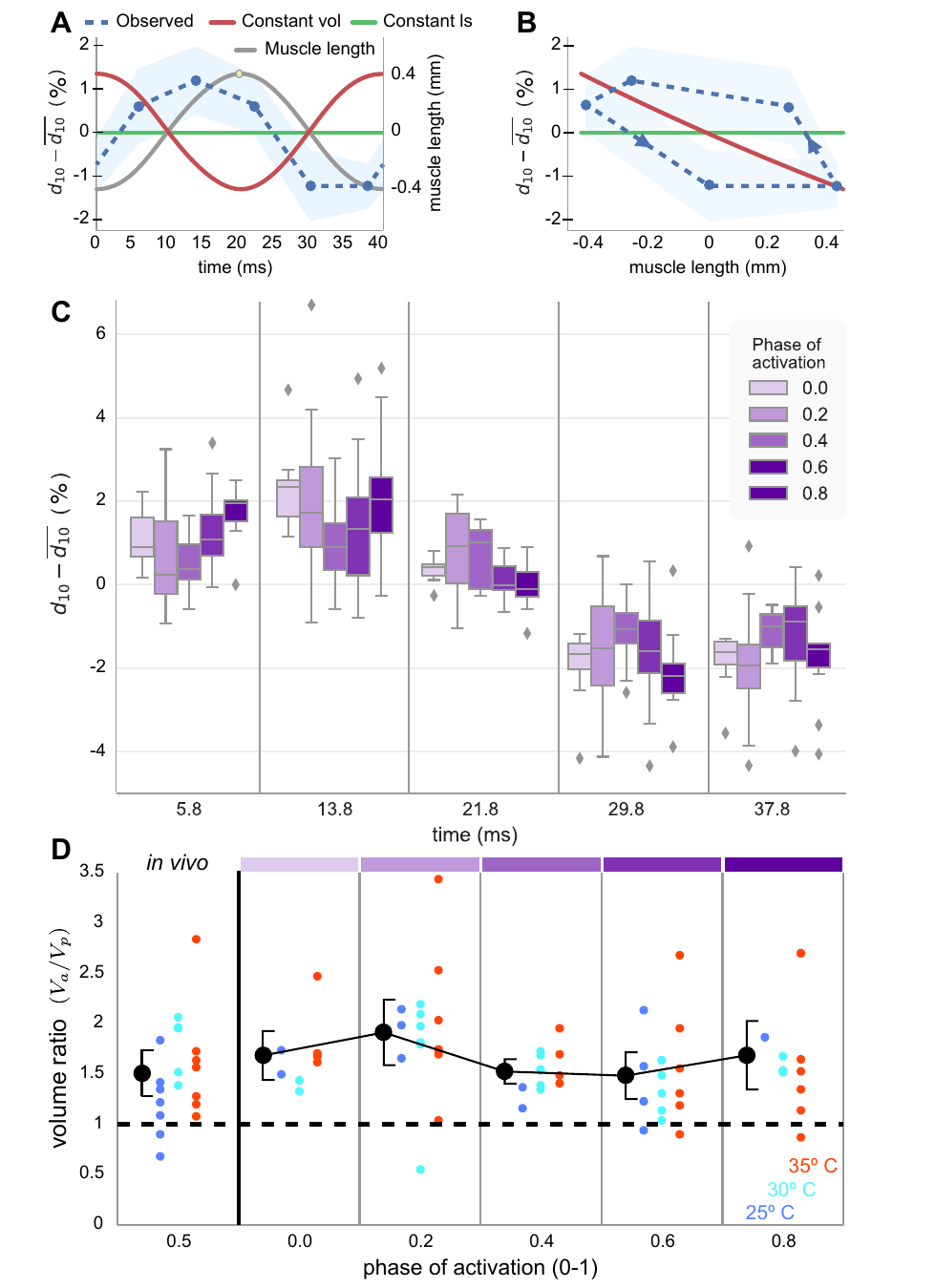}
\internallinenumbers
\caption{ \textbf{Example data measuring the radial lattice spacing.} 
(A) Time-dependent difference between instantaneous and mean $d_{10}$ spacings for muscles experiencing sinusoidal length changes (grey line) at 25~Hz with amplitude of 0.4~mm (strain 4\%) and an activation phase of 0.5 (yellow dot -- onset of shortening). The observed data (blue with $95\%$ CI) differs from the predictions for constant lattice spacing (green) and constant volume (red). (B) The lattice spacing and predictions from A are plotted against the muscle length, illustrating the hysteresis in lattice spacing. (C) Whisker plots (mean, 50\%, 90\% quantiles) for lattice spacing as function of time within the 40 ms length change cycle. The extent and timing of lattice spacing changes depend on the phase of stimulation scaled to the range of 0 to 1 (shades of purple). A phase of 0 corresponds to activation occurring at the onset of lengthening which would be at the beginning of the strain cycle (time = 0~ms). (D) Volume ratio as a function of the phase of activation. The actual volume change of the myofilament lattice during one contraction cycle ($V_a$) is a sum over the time intervals from (C). $V_p$ is what the total volume change would be if the lattice spacing was constant (i.e.\ purely sliding filaments, green line in A \& B). An isovolumetric lattice would produce a value of zero. A ratio of $V_a$ to $V_p$ greater than zero indicates significant lattice volume change and hence advective flow. Ratios greater than 1 indicate even more volume change than would occur if the actin-myosin lattice had constant spacing. Colored data points correspond to different temperatures, black markers are means across all temperatures with 95\% confidence intervals of the mean in black brackets. The \textit{in vivo} condition of a phase of 0.5 is plotted separate from the modified phases which produce systematic variation in the mean volume change, but always result in a volume ratio greater than 1. Color bars at the top show which phases correspond to data in (C).}
\label{fig:fig4}
\end{figure*}

Our time-resolved lattice spacing reveals that \textit{M. sexta} flight muscle does not follow the predictions for constant volume nor for constant lattice spacing. Under typical locomotor conditions ($\sim 10\%$ axial strain, phase of activation $= 0.5$), lattice spacing changes by approximately 2.5\%. With a typical $d_{10}$ spacing of 49 nm, this corresponds to cyclic fluctuations in lattice spacing of 1.2 nm in the radial direction (Fig. \ref{fig:fig4}A). While $\sim 2.5\%$ is comparable to the magnitude of spacing change predicted by the isovolumetric case, the change is not in precise antiphase with axial length change (\ref{fig:fig4}A, B, compare blue to red). As such, there are appreciable volume changes of the lattice.

We quantify the radial lattice spacing dynamics by an effective Poisson ratio, defined as the ratio of lattice spacing change to axial length change. Rather than having a constant effective Poisson ratio of zero (constant spacing - Fig. \ref{fig:fig4}B, green) or 0.5 (constant volume - Fig. \ref{fig:fig4}B, red), the relationship between lattice spacing and axial length is not constant (Fig. \ref{fig:fig4}B, blue). This means that the muscle's lattice has a time-varying effective Poisson ratio. At some points in the length-change cycle, the effective Poisson ratio is positive, at others it is approximately zero, and at others it is negative (behaving briefly as an auxetic material). An effective negative Poisson ratio means that at those times the muscle is shortening axially at the same time that the lattice spacing is being reduced. This could occur when a large number of crossbridges cause both axial and radial contraction. Recent theoretical and experimental evidence suggests that even changes in lattice spacing on the order of $1$ nm can profoundly influence tension development \cite{Williams:2013aa} and rates of cross-bridge attachment or detachment \cite{Williams:2010aa,Williams:2012aa}.

We next altered the phase of muscle stimulation to test the idea that cross-bridge activity influences lattice spacing even when the periodic axial strain remains the same. The synchronous downstroke flight muscle of \textit{M. sexta} receives only a single motor impulse per wingstroke \cite{Tu2004submax}. We altered the timing of cross-bridge force development by changing the phase of this electrical stimulation, defined as the timing of the stimulated spike during the cyclical contractions. The actual activation of thin filaments (related to change in [Ca$^{2+}$]) is not solely dependent on the phase of stimulation, so the effect of changing when the electrical spike occurs likely also alters the magnitude and time course of activation. Regardless of these secondary effects, the timing of peak activation (peak [Ca$^{2+}$]) will change as we alter the phase of electrical stimulation.

Over all phases of stimulation, the radial spacing of the filament lattice varies considerably throughout the length-change cycle (Fig. \ref{fig:fig4}C). In some instances, the peak-to-peak amplitude was as much as $\sim 4.5\%$ (nearly twice that of the activation phase of 0.5 in Fig. \ref{fig:fig4}A, B), while in others the lattice was more constrained (e.g. phase of 0.4). In natural conditions, the stimulation phase is known to vary under neural control \cite{Sponberg2012}. Varying the phase of stimulation results in different lattice dynamics even under the same longitudinal strain.

\subsection*{Change in lattice spacing during contraction enhances volume change beyond predictions even for constant lattice spacing} 
While constant lattice spacing necessarily leads to volume changes of the sarcomere, the magnitude of such changes could be even more extreme because of the time-varying lattice spacing: cross-bridges could pull the myofilaments radially inward as the sarcomere shortens. Using the known axial strain and the measured lattice spacing, we reconstructed the time-varying lattice volume (Fig. \ref{fig:fig4}D). The observed dynamics show 51 +/- 22 \% (95\% CI of mean) more volume change than what would occur for constant lattice spacing (comparing volume ratio to 1, all conditions: $p < 10^{-17},\:N = 85$; \textit{in vivo} conditions:  $p < 10^{-3},\:N = 19$; 35$^\circ$ C alone:  $p = 0.03,\:N = 7$). Temperature can also affect lattice spacing \cite{George:2013aa}, but we varied temperature across the \textit{in vivo} range of 25-35$^\circ$ C and found no significant effect on volume change (Kruskal-Wallis, $p = 0.38,\:N = 85$). The phase of activation did modulate the volume change by $\pm22\%$ (Kruskal-Wallis, $p = 0.03,\:N = 85$), but at all phases volume change was still larger than expected for a constant lattice ($p < 0.01$ in all cases). 

Overall, periodic changes in lattice spacing enhance volume changes in the sarcomeric lattice of \textit{M. sexta} synchronous flight muscle. The degree of lattice spacing and volume change depends partially on the phase of stimulation. This makes sense because crossbridges can actively restrict radial motions and could even enable the simultaneous shortening in both the radial and axial dimensions of the lattice (the effective negative Poisson ratio). This behavior further enhances volume change over the constant lattice spacing conditions. Over long sustained contractions or many cycles, the total volume of the muscle cell can change (e.g., \cite{Neering1991}), even heterogeneously \cite{Rapp1998}, but within each contraction our modeling results show that the experimentally observed volume change results in significant advective flow within the muscle cell while the myofilament lattice changes volume relative to the whole cell.

If lattice spacing dynamics were fully determined by cross-bridge binding, we might expect phases of stimulation that both increase or decrease the volume change compared to the constant lattice spacing condition. However, lattice spacing arises from the interaction of several forces. In addition, the passive, equilibrium lattice spacing might be either greater or less than the steady-state, activated lattice spacing \cite{Tune2020}. This is seen in two very similar cockroach leg muscles, one of which does not change lattice spacing upon isometric activation, but the other of which expands radially \cite{Tune2020}. In contrast, as described above, there are some phases of the length cycle where \textit{M. sexta} flight  muscle radially contracts at the same time as axial contraction. Regardless of direction, the magnitude of the radial component of crossbridges can be comparable to the axial force \cite{Williams:2010aa}. Titin, or in insects the titin-like proteins projectin, kettin, and salimus \cite{Yuan2015}, can also contribute to radial and axial forces and stiffness. Moreover the equilibrium lattice spacing may change when titin-like proteins are activated by Ca$^{2+}$. These proteins may play an underappreciated role both in active force production and in determining lattice spacing \cite{Fukuda2005,Irving2011}. Other structural proteins like myomesin and myosin binding protein C could also contribute to both the radial equilibrium spacing and lattice spacing dynamics during contraction \cite{Schoenauer2005,Razumova2008}. Electrostatic interactions, including Coulomb and van der Waals forces occur between various muscle proteins. With flow occurring in the sarcomere, radial shearing of the filaments could also be significant. Taken together, there are many factors that can influence both the steady-state and dynamic lattice spacing of the sarcomere, so we should not be surprised that different patterns occur in different muscles. At all phases of stimulation, cross-bridge activation in \textit{M. sexta} flight muscle extends for nearly the full contraction cycle \cite{Tu2004submax,Malingen2020}, so it is also possible that the actual volume change ($V_a$) is greater than predicted ($V_p$) at all phases because of work-dependent activation and other dynamic effects \cite{Josephson1999}.

Our experimental results apply for \textit{M. sexta} flight muscle, a synchronous flight muscle with high power demands and a contraction frequency of 25 Hz \cite{Tu2004submax}. The amount of advective advantage due to fluid pumping will certainly depend on the particular pattern of volume change a sarcomere undergoes although it will be present in all case where the sarcomere is not exactly isovolumetric. Other muscles have different patterns of lattice spacing change. In contrast to the patterns seen here, frog muscle's myofilament radially expands during contraction \cite{Cecchi:1990aa}, but while this is closer to the isovolumetric condition, it would still experience some volume changes and an advective advantage, if not exactly isovolumetric. Lattice spacing also decreases during activation under isometric conditions \cite{Bagni1994}, suggesting the potential for advective pumping even when muscle length is constant. \textit{Drosophila} asynchronous flight muscle produces power primarily through delayed stretch activation rather than cyclical Ca$^{2+}$ release and reuptake \cite{Irving2000,Dickinson:2005aa}. Its axial strain is smaller than the 8-10\% peak-to-peak strains in synchronous flight muscle, and it does not show significant lattice spacing change. This would still result in an advective advantage and even though the volume change might be lower in magnitude, the frequencies are an order of magnitude higher. So the magnitude of advective flow likely varies with muscle type, but advective flow is likely a general property that has not been widely considered in sarcomeres.

This advective advantage can contribute to the pumping of metabolites into and out of the sarcomeric lattice beyond what is possible with diffusion alone. Pumping could assist in ATP delivery but could also affect muscle regulation by influencing [Ca$^{2+}$] in the lattice. It could also enhance clearing inorganic phosphate from the myofilament lattice influencing muscle fatigue. Clearing high [P$_i$] could reduce its interference with power stroke kinetics. However, improved P$_i$ transport to the rest of the cell may not be always be beneficial because uptake of P$_i$ by the SR binds Ca$^{2+}$, and hence limits calcium release contributing to fatigue \cite{Nosek1987,Allen2008}. Contraction-induced flows also could affect other aspects of muscle metabolism and excitation-contraction coupling, such as through volume changes in the t-tubule network \cite{McNary2012}. 

\section*{Conclusion}

While muscle cells are likely isovolumetric on short contraction time scales, sarcomeres may vary in volume. Changing sarcomere volume necessitates flow into and out of the myofilament lattice and our modeling shows that these flows can deliver a significant advective advantage over passive diffusion of ATP and other metabolites. Using time-resolved x-ray diffraction, our experiments with insect flight muscle show that this muscle's sarcomeres change volume to a greater extent than if myofilament radial spacing was constant. As a result, there is significant intracellular flow and associated advective transport. Like breathing, myofilament volume changes during contraction and the resulting intra-sarcomeric flow interacts with diffusion to augment substrate exchange. While the exact magnitude depends on the type of muscle and the dynamics of contraction, advective flow due to muscle contraction is a previously unrecognized mechanism that can influence energy delivery, substrate exchange and the flux of regulatory molecules in the crowded intracellular environment of the myofilament lattice.

\section*{Author Contributions}

TLD and SNS designed experiments.  TLD, TCI, and SNS conducted the experiments. CDW extracted data from imaging. CDW, JAC, and SNS analyzed experimental data. JAC, EL, and TLD performed and analyzed the models. All contributed to the writing. 

\section*{Acknowledgments}
We thank Nicole George and Andrew Mountcastle for their illustrations. This project was supported by grant W911NF-14-1-0396 from the Army Research Office to TLD and SNS, National Science Foundation CAREER 1554790 to SNS, grant 9 P41 GM103622 from the National Institute of General Medical Sciences of the National Institutes of Health, and the Richard Komen Endowed Chair to TLD and NSF Physics of Living Systems Student Research Network grant 1205878. This project has also received funding from the European Research Council (ERC) under the European Union's Horizon 2020 research and innovation programme (grant agreement 682754 to EL). This research used resources of the Advanced Photon Source, a U.S. Department of Energy (DOE) Office of Science User Facility operated for the DOE Office of Science by Argonne National Laboratory under Contract No. DE-AC02-06CH11357.

\bibliography{SarcoBreathe}

\begin{thebibliography}{46}
\providecommand{\url}[1]{\texttt{#1}}
\providecommand{\urlprefix}{ }

\bibitem[Kushmerick and Podolsky(1969)]{Kushmerick1969}
Kushmerick, M.~J., and R.~J. Podolsky, 1969.
\newblock {Ionic mobility in muscle cells.}
\newblock \emph{Science} 166:1297--1298.

\bibitem[Carlson et~al.(2014)Carlson, Vigoreaux, and Maughan]{Carlson:2014aa}
Carlson, B.~E., J.~O. Vigoreaux, and D.~W. Maughan, 2014.
\newblock Diffusion coefficients of endogenous cytosolic proteins from rabbit
  skinned muscle fibers.
\newblock \emph{Biophysical Journal} 106:780--792.

\bibitem[Neering et~al.(1991)Neering, Quesenberry, Morris, and
  Taylor]{Neering1991}
Neering, I.~R., L.~A. Quesenberry, V.~A. Morris, and S.~R. Taylor, 1991.
\newblock {Nonuniform volume changes during muscle contraction}.
\newblock \emph{Biophysical Journal} 59:926--933.

\bibitem[Rapp et~al.(1998)Rapp, Ashley, Bagni, Griffiths, and Cecchi]{Rapp1998}
Rapp, G., C.~C. Ashley, M.~A. Bagni, P.~J. Griffiths, and G.~Cecchi, 1998.
\newblock {Volume changes of the myosin lattice resulting from repetitive
  stimulation of single muscle fibers}.
\newblock \emph{Biophysical Journal} 75:2984--2995.

\bibitem[Maughan and Vigoreaux(1999)]{Maughan1999}
Maughan, D.~W., and J.~O. Vigoreaux, 1999.
\newblock {An integrated view of insect flight muscle: genes, motor molecules,
  and motion}.
\newblock \emph{Physiology} 14:87--92.

\bibitem[Askew and Marsh(2001)]{Askew2001}
Askew, G.~N., and R.~L. Marsh, 2001.
\newblock {The mechanical power output of the pectoralis muscle of
  blue-breasted quail (\textit{Coturnix chinensis}): the in vivo length cycle
  and its implications for muscle performance.}
\newblock \emph{Journal of Experimental Biology} 204:3587--3600.

\bibitem[Syme and Josephson(2002)]{Syme2002}
Syme, D.~A., and R.~K. Josephson, 2002.
\newblock {How to build fast muscles: synchronous and asynchronous designs.}
\newblock \emph{Integrative and Comparative Biology} 42:762--770.

\bibitem[Tu and Daniel(2004{\natexlab{a}})]{Tu2004cardiac}
Tu, M.~S., and T.~L. Daniel, 2004.
\newblock {Cardiac-like behavior of an insect flight muscle.}
\newblock \emph{Journal of Experimental Biology} 207:2455--2464.

\bibitem[Girgenrath and Marsh(1999)]{Girgenrath1999}
Girgenrath, M., and R.~L. Marsh, 1999.
\newblock {Power output of sound-producing muscles in the tree frogs
  \textit{Hyla versicolor} and \textit{Hyla chrysoscelis}}.
\newblock \emph{Journal of Experimental Biology} 202:3225--3237.

\bibitem[Rome(2005)]{Rome2005}
Rome, L.~C., 2005.
\newblock {Design amd Function of Superfast Muscles: New Insights into the
  Physiology of Skeletal Muscle}.
\newblock \emph{Annual Review of Physiology} 68:193--221.

\bibitem[Powers et~al.(2018)Powers, Williams, Regnier, and Daniel]{Powers2018}
Powers, J.~D., C.~D. Williams, M.~Regnier, and T.~L. Daniel, 2018.
\newblock {A spatially explicit model shows how titin stiffness modulates
  muscle mechanics and energetics}.
\newblock \emph{Integrative and Comparative Biology} 58:186--193.

\bibitem[Gordon et~al.(2000)Gordon, Homsher, and Regnier]{Gordon2000}
Gordon, A.~M., E.~Homsher, and M.~Regnier, 2000.
\newblock Regulation of contraction in striated muscle.
\newblock \emph{Physiological Reviews} 80:853--924.
\newblock PMID: 10747208.

\bibitem[Irving(2007)]{Irving2007}
Irving, T.~C., 2007.
\newblock {X-ray diffraction of indirect flight muscle from \textit{Drosophila
  in vivo}}.
\newblock \emph{In} J.~O. Vigoreaux, editor, Nature's versatile engine: Insect
  flight muscle inside and out, Springer, New York, 197--213.

\bibitem[Cecchi et~al.(1990)Cecchi, Bagni, Griffiths, Ashley, and
  Maeda]{Cecchi:1990aa}
Cecchi, G., M.~A. Bagni, P.~J. Griffiths, C.~C. Ashley, and Y.~Maeda, 1990.
\newblock Detection of radial crossbridge force by lattice spacing changes in
  intact single muscle fibers.
\newblock \emph{Science} 250:1409--1411.

\bibitem[Cecchi et~al.(1991)Cecchi, Griffiths, Bagni, Ashley, and
  Maeda]{Cecchi:1991aa}
Cecchi, G., P.~J. Griffiths, M.~A. Bagni, C.~C. Ashley, and Y.~Maeda, 1991.
\newblock Time-resolved changes in equatorial x-ray diffraction and stiffness
  during rise of tetanic tension in intact length-clamped single muscle fibers.
\newblock \emph{Biophysical Journal} 59:1273--1283.

\bibitem[Iwamoto and Yagi(2013)]{Iwamoto2013}
Iwamoto, H., and N.~Yagi, 2013.
\newblock {The molecular trigger for high-speed wing beats in a bee}.
\newblock \emph{Science} 341:1243--1247.

\bibitem[George et~al.(2013)George, Irving, Williams, and
  Daniel]{George:2013aa}
George, N.~T., T.~C. Irving, C.~D. Williams, and T.~L. Daniel, 2013.
\newblock The cross-bridge spring: can cool muscles store elastic energy?
\newblock \emph{Science} 340:1217--1220.

\bibitem[Smith(2014)]{Smith2014}
Smith, D.~A., 2014.
\newblock {Electrostatic forces or structural scaffolding: What stabilizes the
  lattice spacing of relaxed skinned muscle fibers?}
\newblock \emph{Journal of Theoretical Biology} 355:53--60.

\bibitem[Williams et~al.(2010)Williams, Regnier, and Daniel]{Williams:2010aa}
Williams, C.~D., M.~Regnier, and T.~L. Daniel, 2010.
\newblock Axial and radial forces of cross-bridges depend on lattice spacing.
\newblock \emph{PLoS Computational Biology} 6:e1001018.

\bibitem[Williams et~al.(2012)Williams, Regnier, and Daniel]{Williams:2012aa}
Williams, C.~D., M.~Regnier, and T.~L. Daniel, 2012.
\newblock Elastic energy storage and radial forces in the myofilament lattice
  depend on sarcomere length.
\newblock \emph{PLoS Computational Biology} 8:e1002770.

\bibitem[Dickinson et~al.(2005)Dickinson, Farman, Frye, Bekyarova, Gore,
  Maughan, and Irving]{Dickinson:2005aa}
Dickinson, M., G.~Farman, M.~Frye, T.~Bekyarova, D.~Gore, D.~Maughan, and
  T.~Irving, 2005.
\newblock Molecular dynamics of cyclically contracting insect flight muscle in
  vivo.
\newblock \emph{Nature} 433:330--334.

\bibitem[Mijailovich et~al.(2017)Mijailovich, Nedic, Svicevic, Stojanovic,
  Walklate, Ujfalusi, and Geeves]{Mijailovich2017}
Mijailovich, S.~M., D.~Nedic, M.~Svicevic, B.~Stojanovic, J.~Walklate,
  Z.~Ujfalusi, and M.~A. Geeves, 2017.
\newblock {Modeling the Actin.myosin ATPase Cross-Bridge Cycle for Skeletal and
  Cardiac Muscle Myosin Isoforms}.
\newblock \emph{Biophysical Journal} 112:984--996.

\bibitem[Oliphant(2006)]{Oliphant:2006aa}
Oliphant, T.~E., 2006.
\newblock A guide to NumPy.
\newblock Trelgol Publishing.

\bibitem[Tu and Daniel(2004{\natexlab{b}})]{Tu2004submax}
Tu, M.~S., and T.~L. Daniel, 2004.
\newblock {Submaximal power output from the dorsolongitudinal flight muscles of
  the hawkmoth Manduca sexta}.
\newblock \emph{Journal of Experimental Biology} 207:4651--4662.

\bibitem[Williams et~al.(2016)Williams, Balazinska, and Daniel]{Williams:2016}
Williams, C.~D., M.~Balazinska, and T.~Daniel, 2016.
\newblock Automated analysis of muscle x-ray diffraction imaging with MCMC.
\newblock \emph{In} F.~Wang, G.~Luo, C.~Weng, A.~Khan, P.~Mitra, and C.~Yu,
  editors, Biomedical Data Management and Graph Online Querying, Springer
  International Publishing, Switzerland, 126--133.

\bibitem[Josephson(1985)]{Josephson1985}
Josephson, R.~K., 1985.
\newblock Mechanical Power output from Striated Muscle during Cyclic
  Contraction.
\newblock \emph{Journal of Experimental Biology} 114:493--512.

\bibitem[Willingham et~al.(2020)Willingham, Kim, Lindberg, Bleck, and
  Glancy]{Willingham2020}
Willingham, T.~B., Y.~Kim, E.~Lindberg, C.~K. Bleck, and B.~Glancy, 2020.
\newblock {The unified myofibrillar matrix for force generation in muscle}.
\newblock \emph{Nature Communications} 11:1--10.

\bibitem[Malingen et~al.(2021)Malingen, Hood, Lauga, Hosoi, and
  Daniel]{Malingen2021}
Malingen, S.~A., K.~Hood, E.~Lauga, A.~Hosoi, and T.~L. Daniel, 2021.
\newblock Fluid flow in the sarcomere.
\newblock \emph{Archives of Biochemistry and Biophysics} 706:108923.

\bibitem[Huxley(1953)]{Huxley:1953aa}
Huxley, H.~E., 1953.
\newblock X-ray analysis and the problem of muscle.
\newblock \emph{Proceedings Of The Royal Society B-Biological Sciences}
  141:59--62.

\bibitem[Matsubara and Elliott(1972)]{Matsubara:1972aa}
Matsubara, I., and G.~F. Elliott, 1972.
\newblock X-ray diffraction studies on skinned single fibres of frog skeletal
  muscle.
\newblock \emph{Journal of Molecular Biology} 72:657--669.

\bibitem[Irving and Maughan(2000)]{Irving:2000aa}
Irving, T.~C., and D.~W. Maughan, 2000.
\newblock \textit{In vivo} x-ray diffraction of indirect flight muscle from
  \textit{Drosophila melanogaster}.
\newblock \emph{Biophysical Journal} 78:2511--2515.

\bibitem[Malingen et~al.(2020)Malingen, Asencio, Cass, Ma, Irving, and
  Daniel]{Malingen2020}
Malingen, S.~A., A.~M. Asencio, J.~A. Cass, W.~Ma, T.~C. Irving, and T.~L.
  Daniel, 2020.
\newblock {In vivo X-ray diffraction and simultaneous EMG reveal the time
  course of myofilament lattice dilation and filament stretch}.
\newblock \emph{Journal of Experimental Biology} 223:jeb224188.

\bibitem[Sponberg and Daniel(2012)]{Sponberg2012}
Sponberg, S., and T.~L. Daniel, 2012.
\newblock {Abdicating power for control: a precision timing strategy to
  modulate function of flight power muscles.}
\newblock \emph{Proceedings Of The Royal Society B-Biological Sciences}
  279:3958--3966.

\bibitem[Williams et~al.(2013)Williams, Salcedo, Irving, Regnier, and
  Daniel]{Williams:2013aa}
Williams, C.~D., M.~K. Salcedo, T.~C. Irving, M.~Regnier, and T.~L. Daniel,
  2013.
\newblock The length-tension curve in muscle depends on lattice spacing.
\newblock \emph{Proceedings Of The Royal Society B-Biological Sciences}
  280:20130697.

\bibitem[Tune et~al.(2020)Tune, Ma, Irving, and Sponberg]{Tune2020}
Tune, T.~C., W.~Ma, T.~Irving, and S.~Sponberg, 2020.
\newblock {Nanometer-scale structure differences in the myofilament lattice
  spacing of two cockroach leg muscles correspond to their different
  functions}.
\newblock \emph{The Journal of Experimental Biology} 223:jeb212829.

\bibitem[Yuan et~al.(2015)Yuan, Ma, Schemmel, Cheng, Liu, Tsaprailis, Feldman,
  {Ayme Southgate}, and Irving]{Yuan2015}
Yuan, C.-C., W.~Ma, P.~Schemmel, Y.-S. Cheng, J.~Liu, G.~Tsaprailis,
  S.~Feldman, A.~{Ayme Southgate}, and T.~C. Irving, 2015.
\newblock {Elastic proteins in the flight muscle of \textit{Manduca sexta}}.
\newblock \emph{Archives of Biochemistry and Biophysics} 568:16--27.

\bibitem[Fukuda et~al.(2005)Fukuda, Wu, Farman, Irving, and
  Granzier]{Fukuda2005}
Fukuda, N., Y.~Wu, G.~Farman, T.~C. Irving, and H.~Granzier, 2005.
\newblock {Titin-based modulation of active tension and interfilament lattice
  spacing in skinned rat cardiac muscle}.
\newblock \emph{Pfl\"{u}gers Archiv - European Journal of Physiology}
  449:449--457.

\bibitem[Irving et~al.(2011)Irving, Wu, Bekyarova, Farman, Fukuda, and
  Granzier]{Irving2011}
Irving, T., Y.~Wu, T.~Bekyarova, G.~P. Farman, N.~Fukuda, and H.~Granzier,
  2011.
\newblock {Thick-filament strain and interfilament spacing in passive muscle:
  Effect of titin-based passive tension}.
\newblock \emph{Biophysical Journal} 100:1499--1508.

\bibitem[Schoenauer et~al.(2005)Schoenauer, Bertoncini, Machaidze, Aebi,
  Perriard, Hegner, and Agarkova]{Schoenauer2005}
Schoenauer, R., P.~Bertoncini, G.~Machaidze, U.~Aebi, J.-C. Perriard,
  M.~Hegner, and I.~Agarkova, 2005.
\newblock {Myomesin is a Molecular Spring with Adaptable Elasticity}.
\newblock \emph{Journal of Molecular Biology} 349:367--379.

\bibitem[Razumova et~al.(2008)Razumova, Bezold, Tu, Regnier, and
  Harris]{Razumova2008}
Razumova, M.~V., K.~L. Bezold, A.-Y. Tu, M.~Regnier, and S.~P. Harris, 2008.
\newblock {Contribution of the Myosin Binding Protein C Motif to Functional
  Effects in Permeabilized Rat Trabeculae}.
\newblock \emph{Journal of General Physiology} 132:575--585.

\bibitem[Josephson(1999)]{Josephson1999}
Josephson, R.~K., 1999.
\newblock {Dissecting muscle power output}.
\newblock \emph{Journal of Experimental Biology} 202:3369--3375.

\bibitem[Bagni et~al.(1994)Bagni, Cecchi, Griffiths, Ma{\'{e}}da, Rapp, and
  Ashley]{Bagni1994}
Bagni, M., G.~Cecchi, P.~Griffiths, Y.~Ma{\'{e}}da, G.~Rapp, and C.~Ashley,
  1994.
\newblock {Lattice spacing changes accompanying isometric tension development
  in intact single muscle fibers}.
\newblock \emph{Biophysical Journal} 67:1965--1975.

\bibitem[Irving et~al.(2000)Irving, Konhilas, Perry, Fischetti, and
  de~Tombe]{Irving2000}
Irving, T.~C., J.~Konhilas, D.~Perry, R.~Fischetti, and P.~P. de~Tombe, 2000.
\newblock {Myofilament lattice spacing as a function of sarcomere length in
  isolated rat myocardium}.
\newblock \emph{American Journal of Physiology-Heart and Circulatory
  Physiology} 279:H2568--H2573.

\bibitem[Nosek et~al.(1987)Nosek, Fender, and Godt]{Nosek1987}
Nosek, T., K.~Fender, and R.~Godt, 1987.
\newblock {It is diprotonated inorganic phosphate that depresses force in
  skinned skeletal muscle fibers}.
\newblock \emph{Science} 236:191--193.

\bibitem[Allen et~al.(2008)Allen, Lamb, and Westerblad]{Allen2008}
Allen, D.~G., G.~D. Lamb, and H.~Westerblad, 2008.
\newblock {Skeletal Muscle Fatigue: Cellular Mechanisms}.
\newblock \emph{Physiological Reviews} 88:287--332.

\bibitem[McNary et~al.(2012)McNary, Spitzer, Holloway, Bridge, Kohl, and
  Sachse]{McNary2012}
McNary, T.~G., K.~W. Spitzer, H.~Holloway, J.~H. Bridge, P.~Kohl, and F.~B.
  Sachse, 2012.
\newblock {Mechanical modulation of the transverse tubular system of
  ventricular cardiomyocytes}.
\newblock \emph{Progress in Biophysics and Molecular Biology} 110:218--225.

\end{thebibliography}


\section*{Supplementary Material}

An online supplement to this article can be found by visiting BJ Online at \url{http://www.biophysj.org}.

\end{document}